\documentclass{aipproc}

\usepackage{epsf}

\layoutstyle{6x9}

\newif\ifpreprint

\preprinttrue

\begin{document}

\title{
\ifpreprint
\rightline{\normalsize UASLP--IF--02--002}
\vspace{0.5cm}
\fi
Resonances in 
\ifpreprint
$\Lambda_c^+ \to pK^-\pi^+$\thanks{Talk given
at the VIII Mexican Workshop on Particles and Fields, November 14-20, 2001,
Zacatecas, Mexico. Proceedings to be published by AIP.}
\else
$\Lambda_c^+ \to pK^-\pi^+$
\fi
}
\author{
\centerline{
\ifpreprint
Juan Medellin Z.\thanks{email: {\tt juan@ifisica.uaslp.mx}}, 
J\"urgen Engelfried\thanks{email: {\tt jurgen@ifisica.uaslp.mx}},
Antonio Morelos\thanks{email: {\tt morelos@ifisica.uaslp.mx}}
\else
Juan Medellin Z., 
J\"urgen Engelfried,
Antonio Morelos
\fi
}
\centerline{For the SELEX Collaboration}
}{
address={
\centerline{
Instituto de F\'{\i}sica, Universidad Aut\'onoma de San Luis Potos\'{\i},}
\centerline{
\'Alvaro Obregon 64, Zona Centro, San Luis Potos\'{\i}, S.L.P.~78000, M\'exico}
}}

\begin{abstract}
We report very preliminary results of a Dalitz-plot analysis~\cite{tesis}
 of $\Lambda_c^+$ in
the decay to $p K^-\pi^+$ with the  helicity formalism.
We used the data from the fixed 
target experiment SELEX~\cite{SELEX} (E781) in Fermilab. 
We report about branching-ratios
of the resonant states involved,  and a possible initial state polarization.
\end{abstract}

\maketitle

\section{Introduction and Analysis Method}

In our search for resonances in the 3-body decay of
$\Lambda_c^+\to p K^-\pi^+$ we are considering in addition to
the non-resonant mode the following resonant decay modes:
$\Lambda_c^+ \to \bar{K^{*0}} p$, $\bar{K^{*0}}  \to K^- \pi^+ $;
$\Lambda_c^+ \to \Delta^{++} K^-$, $\Delta^{++} \to p \pi^+$; 
and
$\Lambda_c^+ \to  \Lambda(1520) \pi^+$, $\Lambda(1520) \to p K^-$.

\begin{figure}[htb]
\leavevmode
\epsfxsize=\hsize
\epsffile{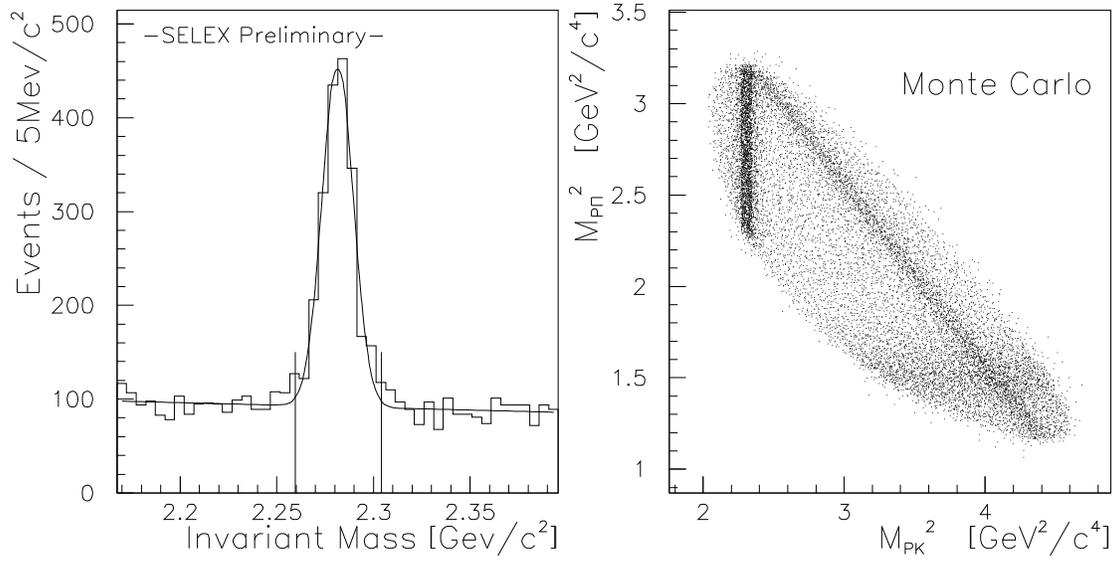}
\caption{Left: Invariant mass distribution for $p K^-\pi^+$. A signal of
approximately 1500~$\Lambda_c^+$ can be clearly seen.  The signal
and the sideband regions used in this analysis are indicated.
Right: Dalitz plot from Monte Carlo with a mixture of non-resonant and 
three resonant decay modes.}
\label{lambdac}
\end{figure}

As a search tool for the resonances we calculate the invariant masses
of pairs of daughter particles $M^{2}_{ij}$, $M^{2}_{ik}$
and fill with them into a two-dimensional histogram  (Dalitz plot).
A Monte-Carlo simulation of what could be expected is shown 
in fig.\,\ref{lambdac}~(right).

The data sample used for this analysis is the same as in a previous
SELEX publication~\cite{lambdaclife}, where more details
can be found. We applied the following
cuts to get a clean $\Lambda_c^+$ signal:
good fits for tracks and vertices;
momentum $>8.0\,\mbox{GeV}/c$ for all tracks;
proton and kaon identified in the RICH~\cite{selexrich};
secondary vertex outside material;
separation~$L$ of primary and secondary vertices $L>8\,\sigma$,
where $\sigma$ denotes the combined error of the vertices;
$\sigma<1.7\,\mbox{mm}$;
at least two tracks with a miss distance to the primary 
vertex of more than $20\,\mu\mbox{m}$;
the momentum vector of the reconstructed $\Lambda_c^+$ has to point back
to the primary vertex. 
The obtained $p K^-\pi^+$ invariant mass distribution can be found in
fig.\,\ref{lambdac}~(left).

\begin{figure}[htb]
\leavevmode
\epsfxsize=\hsize
\epsffile{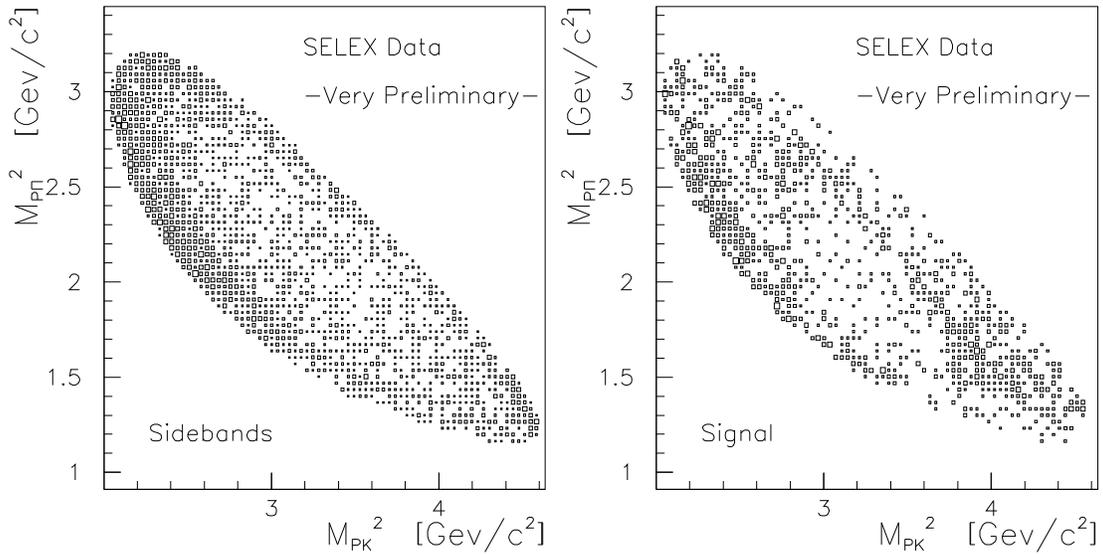}
\caption{Left: Dalitz plot for the sideband region indicated in
fig.\,\ref{lambdac} with phase space mapping.
Right: Dalitz plot of signal region after sideband subtraction.}
\label{dalitz}
\end{figure}

To eliminate the contribution of background under the $\Lambda_c^+$-peak,
we produce a Dalitz-plot of the sidebands shown in fig.\,\ref{lambdac},
with a proper mapping of the allowed phase space.   After normalizing
to the correct number of events, we subtract this contribution bin-by-bin
from a Dalitz plot of the signal region. 
We verified this procedure by comparing the plots obtained by different
sideband regions on the left and right side of the signal region.
In fig.\,\ref{dalitz} we show
the Dalitz plots for the sideband and the signal regions.
In the background subtracted signal region one can still see some enhancement
on the lower left of the phase space, also present in the
sideband; we are still investigating the nature of this.

To extract the information from the Dalitz plot, we used an helicity
formalism~\cite{fox,brasil}. We will describe the fit functions used
in the following sections.

\section{Helicity Formalism for Two Body Decays}
From Fermi's Golden Rule the partial decay width is given by
$d\Gamma=\frac{(2\pi)^4}{2M}\vert \Omega \vert ^2 d\Phi_n$, with
$d\Gamma \sim \vert \Omega \vert ^2 = \vert<\!BC\vert T\vert A\!>\vert^2$,
where $A$ and $BC$ denote the initial and final states.
We are working in the eigenstate base
$\vert M, S, p^{\mu}, \lambda \!>$  of the operators
$M^2 = P_{\mu} P^{\mu}$,
$S = -\omega_{\mu} \omega^{\mu}$ with 
$ \omega^{\sigma} = 
\frac{1}{2} \epsilon^{\sigma \mu \nu \lambda}\ M_{\mu \nu}\ P_{\lambda}$, and
$P^{\mu}, \omega^O = J_x P_x + J_y P_y + J_z P_z$;
$\omega^O$ represents the helicity operator with eigenvalue $\lambda$. 
For simplicity we assume only $\vert  p^{\mu}, \lambda \!>$.

In the rest frame of the mother particle~$A$ we can express the
spin states as $\vert j_A m_A\!>$
and the final state as $\vert \theta_B \phi_B \lambda_B \lambda_C\!>$, with
\begin{displaymath}
\vert \theta_B \phi_B \lambda_B \lambda_C\!> = 
\sum_{JM} \sqrt{\frac{2J+1}{4\pi}} 
D_{M\lambda_1}^J (\phi_B, \theta_B, -\phi_B) \vert JM\lambda_B \lambda_C\!>.
\end{displaymath}
Applying angular momentum conservation, we can write the transition
amplitude as
\begin{displaymath}
<\!BC\vert T\vert A\!> = \sqrt{\frac{2j_A+1}{4\pi}} 
D_{m_A\lambda_1}^{\star j_A}
(\phi_B, \theta_B, -\phi_B) <\!\lambda_B \lambda_C\vert T\vert j_A m_A\!>
\end{displaymath}
With 
$D_{m\lambda}^{j}(\phi, \theta, -\phi) =
e^{-i\phi (m - \lambda)} d_{m \lambda}^j (\theta)$,
$\lambda_1=\lambda_B-\lambda_C$ ($|\lambda_1|\le m_A$), 
and summing
over
all spin and helicity projections and we obtain
\begin{displaymath}
d\Gamma \sim
\sum_{m_A} \sum_{\lambda_B \lambda_C} 
\vert  \alpha_{\lambda_B \lambda_C} 
e^{i\phi_B (m_A - \lambda_1)} d_{m_A \lambda_1}^{j_A} (\theta_B) \vert^2
\end{displaymath}

\section{Helicity Formalism for Three Body Resonant Decays}
The decay width in this case is given by
$d\Gamma  \sim \vert <\!DE\vert T_2\vert B\!> <\!BC\vert T_1\vert A\!>\vert^2$.
In the rest frame of the resonance~$B$, with the $z$-axes pointing
into the direction of motion of the resonance~$B$ in the rest system
of the mother particle~$A$, we can write
\begin{displaymath} 
<\!DE\vert T_2\vert B\!> =
\sqrt{\frac{2j_B+1}{4\pi}} e^{i\phi_D^{'} (\lambda_B - 
\lambda_2)} d_{\lambda_B \lambda_2}^{j_B} (\theta_D^{'}) 
<\!\lambda_D \lambda_E\vert T_2\vert B\!>.
\end{displaymath}
Summing over all resonances we obtain
\begin{equation}
d\Gamma \sim \sum_{m_A} \sum_{\lambda_C \lambda_D \lambda_E} 
\mathbf{ \vert}  \sum_{\lambda_B}  \sum_{B} 
BW(m_r) \alpha_{\lambda_B \lambda_C}  
\alpha_{\lambda_D \lambda_E} e^{i\phi_B (m_A - \lambda_1)} 
e^{i\phi_D^{'} (\lambda_B - \lambda_2)} 
d_{m_A \lambda_1}^{j_A} (\theta_B) d_{\lambda_B \lambda_2}^{j_B} 
(\theta_D^{'})\vert^2
\label{helicity}
\end{equation}
where we also consider the finite width of the resonance with a 
Breit-Wigner distribution
$BW(m_r) \sim \frac{ m_0 \Gamma_r}{m_r^2 - m_0^2 + i m_r \Gamma_r}$.
In the parity conserving  decay of the resonance we can write
$\alpha_{\lambda_D \lambda_E} = (-1)^{S_D+S_E-S_B} \eta_B \eta_D \eta_E
  \alpha_{-\lambda_D -\lambda_E}$.

In this formalism we can naturally introduce an initial polarization
of the mother particle $P_A$, given by
$d\Gamma \sim 
\frac{1}{2} (1+P_A) \sum_{\lambda_C \lambda_D \lambda_E} 
\vert  \sum_{\lambda_B}  \sum_{B} 
BW(m_B)
\xi_{B,\frac{1}{2},\lambda_B,\lambda_C,\lambda_D,\lambda_E} \vert^2 
$ \\ $
+\frac{1}{2} (1-P_A) \sum_{\lambda_C \lambda_D \lambda_E} 
\vert  \sum_{\lambda_B}  \sum_{B} 
BW(m_B) \xi_{B,- \frac{1}{2},\lambda_B,\lambda_C,\lambda_D,\lambda_E} 
\vert^2$.

Integrating the contribution of a resonance over the phase space gives
\begin{equation}
F_r = \frac{\int \sum_{m_A, \lambda_B} \vert 
BW(m_r) \xi_{r,m_A,\lambda_B} \vert^2 
d\vec{x}}{\int \sum_{m_A, \lambda_B} \vert \sum_{B} 
BW(m_B)  \xi_{B,m_A,\lambda_B} \vert^2 d\vec{x}}
\label{fitfraction}
\end{equation}
from where, after applying weight factors for isospin conservation,
we can extract the branching ratio for the resonance.

\section{Preliminary Results}
We performed an unbinned maximum likelihood fit with the functions described
in the previous section, with 23~free parameters in the fit.
The preliminary results are shown in the following tables. 
Only statistical errors are shown.

\begin{minipage}[t]{8.5cm}
\begin{tabular}{|c|c|c|c|}
\hline
&  Parameter & Amplitude &  Phase \\ \hline
&  $N_1$ &1. fixed  &0. fixed  \\
nonres. & $N_2$ & $320\pm 82$ & $.2\pm 1.9$ \\
&  $N_3$ & $26\pm 56$  &  $4.7\pm 1.9$ \\
&  $N_4$ &  $200\pm 135$  & $3.2\pm 0.2$   \\
$\bar K^{*0}$ & $A_1$ &  $495\pm 143$ & $5.6\pm 0.2$   \\
&  $A_2$ &  $70\pm 91$ &  $3.9\pm 10^3$   \\
$\bar K^{*0}$ & $A_3$ &  $362\pm 91$ &  $3.3\pm 0.1$   \\
& $A_4$ & $95\pm 69$ & $2.9\pm 481$   \\
& $B_1$ &  $11\pm 23$ & $3.3\pm 10^4$  \\
$\Delta^{++}$ & $B_2$ &  $196\pm 87$ &  $3.5\pm .9$  \\
& $C_1$ &  $115\pm 142$ & $2.6\pm 18$  \\
$\Lambda(1520)$ & $C_2$ &  $644\pm 144$ &  $0.8\pm .2$  \\
\hline
\end{tabular}
\end{minipage}
\begin{minipage}[t]{5.5cm}
\begin{tabular}{|c|c|}
\hline
Polarization & $P_A=0.1\pm 0.4$\\
\hline
\multicolumn{2}{c}{~~}\\
\hline
\multicolumn{2}{|c|}{Branching Ratio}\\ \hline
 nonres.   & $0.73\pm 0.29$ \\
 $\bar K^{*0}$   & $0.14\pm 0.17$ \\
 $\Delta^{++}$   & $0.09\pm 0.14$ \\
 $\Lambda(1520)$ & $0.04\pm 0.09$ \\
\hline
\end{tabular}
\end{minipage}
The two most significant features in the Dalitz plot are not properly
taken into account in this analysis method. The fit function does
not include a resonance describing the feature seen in the lower left
of the phase space, leading to an over-estimate of the non-resonant
contribution, and the central region with a small number of entries
is not taken into account in an unbinned fit.
%
%
At this moment, we are finalizing the analysis procedure, by
optimizing the cuts used for extracting the $\Lambda_c^+$ signal,
a study if additional resonances have to be included,
and on a binned fit to included the information about the center region.
Also some systematic studies are under way to
quantify the significance of the results.
This work was supported by CONACyT Mexico under Grant~28435-E.

\end{document}